# STABILITY ANALYSIS OF FRACTIONAL RELATIVISTIC POLYTROPES

Mohamed S. Aboueisha[1], A. S. Saad[2,1], Mohamed I. Nouh[1,*], Tarek M. Kamel[1], M. M. Beheary[3] and Kamel A. K. Gadallah[3]

[1] Astronomy Department, National Research Institute of Astronomy and Geophysics, 11421 Helwan, Cairo, Egypt.
[2] Administrative Information System Department, College of Business and Economics, Qassim University, Buraidah 51452, Qassim, Saudi Arabia
[3] Department of Astronomy and Meteorology, Faculty of Science, Al-Azhar University, Nasr City, 11889, Cairo, Egypt.

\* Corresponding author (mohamed.nouh@nriag.sci.eg)

**Abstract:** In astrophysics, the gravitational stability of a self-gravitating polytropic fluid sphere is an intriguing subject, especially when trying to comprehend the genesis and development of celestial bodies like planets and stars. This stability is the sphere's capacity to stay in balance in the face of disruptions. We utilize fractional calculus to explore self-gravitating, hydrostatic spheres governed by a polytropic equation of state $P = K\rho^{1+1/n}$. We focus on structures with polytropic indices ranging from 1 to 3 and consider relativistic and fractional parameters, denoted by $\sigma$ and $\alpha$, respectively. The stability of these relativistic polytropes is evaluated using the critical point method, which is associated with the energetic principles developed in 1964 by Tooper. This approach enables us to pinpoint the critical mass and radius at which polytropic spheres shift from stable to unstable states. The results highlight the critical relativistic parameter ($\sigma_{CR}$) where the polytrope's mass peaks, signaling the onset of radial instability. For polytropic indices of $n = 1$, 1.5, 2, and 3 with a fractional parameter $\alpha = 1$, we observe stable relativistic polytropes for $\sigma$ values below the critical thresholds of $\sigma_{CR} = $ 0.42, 0.20, 0.10, and 0.0, respectively. Conversely, instability emerges as $\sigma$ surpasses these critical values. Our comprehensive calculations reveal that the critical relativistic value ($\sigma_{CR}$) for the onset of instability tends to increase as the fractional parameter $\alpha$ decreases.

## 1. Introduction

The gravitational stability of a self-gravitating polytropic fluid sphere is a fascinating topic in astrophysics, particularly in understanding the formation and evolution of astrophysical objects such as stars and planets. This stability refers to the sphere's ability to maintain equilibrium among disturbances. Analysis typically involves scrutinizing minor deviations from equilibrium and assessing whether these fluctuations amplify or diminish over time. The determination of the stability criteria relies on several factors, such as the polytropic index, total mass, and sphere radius [1]. Compact objects, thought to be composed of a dense fluid of neutrons and other particles, have a polytropic index typically assumed to be between 0.5 and 3.0. Consequently, the stability of relativistic polytropes, specifically within the range of these polytropic indices, is of significant interest, as referenced in previous studies [2-9].

Exploration of stability analyses can catalyze the creation of new theoretical frameworks and the expansion of analytical tools. These models play a pivotal role in enhancing our comprehension of relativistic astrophysics while broadening the resources accessible to researchers. Therefore, stability analysis actively propels the progress of the entire field [10-12].

By solving the TOV equation for a range of central densities and polytropic indices, one can determine the critical relativistic values ($\sigma_{CR}$) beyond which the sphere becomes gravitationally unstable due to relativistic effects. These critical values provide insights into self-gravitating polytropic fluid spheres' maximum mass and stability limits in the relativistic regime. When the relativistic effects $\sigma$, exceed the critical relativistic value $\sigma_{CR}$, it indicates that the self-gravitating polytropic fluid sphere has surpassed a stability limit due to relativistic effects. This implies that the sphere is gravitationally unstable in a relativistic regime, which can have significant consequences for the behavior of the star's quantities. The most immediate consequence is gravitational collapse. When the relativistic effects become dominant and exceed the critical value, the pressure generated by the fluid within the star is insufficient to counteract the inward pull of gravity. This can lead to the collapse of the star's core, resulting in a more compact object, such as a neutron star or even a black hole, depending on the mass of the collapsing core. As the collapse progresses, the density and temperature within the core increase dramatically [10-11]. This can trigger various physical processes, such as nuclear fusion, in the core regions where conditions become extreme. These processes can release vast amounts of energy, leading to violent and energetic events.

Numerous studies have investigated the stability of polytropic models, yielding noteworthy findings. The researcher [13] revealed that self-gravitating polytropic spheres with a polytropic index (n = 3) exhibit unconditional stability against radial perturbations. A pioneering contribution came from [14] when he introduced the radial stability equation. Earlier methodologies for investigating the stability of polytropic stars can be traced in [15] and [16]. Recent advancements have extended our comprehension of polytropic stability with varying polytropic indices, as exemplified by [17-18]. Studies by [19-20] conducted a comprehensive analysis of polytropic sphere stability, employing two distinct approaches, namely the energetic and dynamic methods based on the study of radial pulsations. In a recent study by [21], the critical value of the relativistic parameter for relativistic polytropes, spanning polytropic indices ranging from 1.0 to 3.0, was thoroughly examined. The investigation revealed a noteworthy trend: as the polytropic index increases, the critical value of the relativistic parameter diminishes. For instance, when the polytropic index is n = 1, the critical value stands at σ = 0.42, while for n = 2.5, it notably decreases to σ = 0.04.

In the last two decades, fractional differential equations have seen a surge of interest across various scientific and engineering disciplines, encompassing mathematics, chemistry, optics, plasma physics, and fluid dynamics. Notably, there has been a growing focus on the applications of fractional calculus in astrophysics, as highlighted by [22].

Fractional calculus represents a relatively recent frontier in mathematics, with its potential applications in physics still unfolding. Conducting a stability analysis of fractional relativistic polytropes offers a unique avenue for investigating how fractional calculus might be applied to examining gravitational dynamics and other physical phenomena [23-28]. Fractional relativistic polytropes allow for exploring a wide range of polytropic indices. The objective here is to examine how changes in the polytropic index, a measure of the stiffness of the equation of state, affect stability. Different polytropic indices correspond to varying physical conditions within the object. By studying their effects on stability, researchers gain insights into the diversity of compact astrophysical objects [29-31].

In the present study, we apply the fractional derivatives approach to explore the Tolman-Oppenheimer-Volkoff equation (FTOV), focusing on the stability analysis of relativistic polytropes with a range of polytropic indices. Our approach yields an analytical solution to the FTOV, facilitating the derivation of key physical parameters that define the relativistic polytrope's characteristics. A significant aspect of our research involves determining the relativistic parameter's critical values that signal the onset of radial instability. Working within the modified

Riemann Liouville (mRL, Appendix A1) framework, we rigorously analyze the stability of the fractional relativistic polytrope gas sphere across various polytropic indices. A novel aspect of our work is developing a new analytic solution for the fractional relativistic polytropic gas sphere, achieved through the series expansion method. This solution's foundation is the acceleration of the power series solution of the FTOV, achieved by employing a combination of two distinct transformations. The structure of the paper is as follows: section 2 is devoted to the formulation and solution of the FTOV. Section 3 describes the numerical results. Then, the conclusion is outlined in section 4.

## 2. Basic Equations

In the present section, we shall briefly recall the basic equations implemented to calculate the fractional relativistic polytropic sphere; full details can be found in [31]. In the case of spherically symmetric (r, $\vartheta$, $\varphi$) and static spacetime (t), the line element signature is given by [32]

$$d\,s^2 = e^{2v(r)}\left(cd\,t\right)^2 - e^{2\lambda(r)}d\,r^2 - r^2 d\,\vartheta^2 - r^2 \sin^2\left(\vartheta\right)d\,\varphi^2 \,, \tag{1}$$

Equation (1) can be written in the fractional form as [31]

$$d^\alpha s^2 = e^{2v}\left(cd^\alpha t\right)^2 - e^{2\lambda}d^\alpha r^2 - r^{2\alpha} d^\alpha \vartheta^2 - r^{2\alpha} \sin^2\left(\vartheta^\alpha\right)d^\alpha \varphi^2 \,, \tag{2}$$

where the metric coefficients $v$ and $\lambda$ are functions of $r^\alpha$ only and governed by Einstein's field equation.

If $\rho$ is the mass density and $p$ isotropic pressure in the rest frame of the fluid, the law of local energy-momentum conservation in the fractional form may be written as [31]

$$(\rho c^2 + p)D_r^\alpha v = -D_r^\alpha p \,, \tag{3}$$

which is the fractional hydrostatic equilibrium equation that describes the balance between gravitational force and pressure gradient. The fractional scalar curvature and the fractional Ricci tensor (the (t)(t) component) is given by

$$^\alpha G_{(t)(t)} = \frac{1}{r^{2\alpha}} - \frac{e^{-2\lambda}}{r^{2\alpha}} - \frac{1}{r^\alpha} D_r^\alpha e^{-2\lambda} = \frac{8\pi G}{c^2}\rho \,, \tag{4}$$

$G$ is the gravitational acceleration, $c$ is the light velocity; $v$ and $\lambda$ are functions of $r^\alpha$ only and governed by Einstein's field equation.

Equation (4) can be transferred into the form

$$D_r^\alpha \left( r^\alpha (1-e^{-2\lambda}) \right) = D_r^\alpha \left( \frac{ZG}{c^2} m(r^\alpha) \right), \tag{5}$$

where $Z = \Gamma(2\alpha+1)/\Gamma(\alpha+1)$ and the mass contained in the radius $r^\alpha$ is given by

$$m(r^\alpha) = \int_0^{r^\alpha} 4\pi x^\alpha \rho(x^\alpha) dx^\alpha, \tag{6}$$

from Equation (5), we have the relation

$$e^{2\lambda} = \left[ 1 - \frac{ZG\, m(r^\alpha)}{c^2 R} \right]^{-1}. \tag{7}$$

The (r)(r) component of the field equations reads

$$^\alpha G_{(r)(r)} = -\frac{1}{r^{2\alpha}} + \frac{Ze^{-2\lambda}}{r^\alpha} D_r^\alpha v = \frac{8\pi G}{c^4} p,$$

so, we obtain the relation

$$D_r^\alpha v = \frac{\frac{ZG}{c^2} m(r^\alpha) + \frac{8\pi G}{c^4} pr^{3\alpha}}{Zr^\alpha \left[ r^\alpha - \frac{ZG}{c^2} m(r^\alpha) \right]}. \tag{8}$$

The fractional equation of hydrostatic equilibrium may now be expressed in the fractional Tolman-Oppenheimer-Volko (FTOV) form as

$$D_r^\alpha p = -(\rho c^2 + p) \frac{\frac{ZG}{c^2} m(r^\alpha) + \frac{8\pi G}{c^4} pr^{3\alpha}}{Zr^\alpha \left[ r^\alpha - \frac{2G}{c^2} m(r^\alpha) \right]},$$

or in another form, as

$$D_r^\alpha p = -\frac{GM\rho}{r^{3\alpha}} \left[ 1 + \frac{p}{c^2 \rho} \right] \left[ 1 + \frac{4\pi r^{3\alpha} p}{Zc^2 M} \right] \left[ 1 - \frac{ZGm(r^\alpha)}{c^2 r^\alpha} \right]^{-1}. \tag{9}$$

Equation (9) provides the equilibrium solution for pressure in a compact star when combined with the mass formula, Equation (6), and a microscopic explanation of the link between pressure and energy density. The relationship established by the polytropic equation of state between the fluid's pressure and energy density is given by

$$\rho = \rho_c \theta^n, \tag{10}$$

$$p = K\rho^{1+\frac{1}{n}}, \tag{11}$$

where $n$, $K$, and $\theta$ are the polytropic index, the pressure constant, and the Emden function, respectively.

Inserting Equations (10) and (11) in Equation (3) yields

$$\Gamma(\alpha+1)\frac{K\rho_c^{1+\frac{1}{n}}}{c^2}(1+n)D_r^\alpha \theta + \left(\frac{K\rho_c^{1+\frac{1}{n}}}{c^2}\theta + \rho_c\right)D_r^\alpha v = 0, \tag{12}$$

Now write Equation (12) as

$$\sigma\,\Gamma(\alpha+1)(1+n)D_r^\alpha \theta + (1+\sigma\,\theta)D_r^\alpha v = 0. \tag{13}$$

where the relativistic parameter $\sigma$ is given by

$$\sigma = \frac{K\rho_c^{1/n}}{c^2}.$$

From Equations (4) and (7), we obtain

$$\frac{e^{-2\lambda}}{r^{2\alpha}}(1+Zr^\alpha D_r^\alpha v) - \frac{1}{r^{2\alpha}} = \frac{8\pi G}{c^4}p, \tag{14}$$

$$\frac{1}{r^{2\alpha}}\left(1 - \frac{ZGm(r)}{r^\alpha c^2}\right)(1+Zr^\alpha D_r^\alpha v) - \frac{1}{r^{2\alpha}} = \frac{8\pi G}{c^4}p, \tag{15}$$

inserting Equations (10) and (11) in Equation (15) and put $r = A\xi$, we get

$$\xi^{2\alpha}D_\xi^\alpha \theta - \frac{\sigma(1+n)}{(1+\sigma\,\theta)}\frac{(1+\sigma\,\theta)}{\sigma(1+n)}\frac{AZG\,m}{c^2}\xi^\alpha D_\xi^\alpha \theta - \frac{(1+\sigma\,\theta)}{\sigma(1+n)}\frac{G\,m\,A}{c^2}$$
$$-\frac{(1+\sigma\,\theta)}{\sigma(1+n)}\frac{8\pi G}{A^2 Zc^4}\sigma\,\theta\,\rho\xi^{3\alpha} = 0 \tag{16}$$

where

$$A = \left(\frac{8\pi G \rho_c}{Z\sigma(n+1)c^2}\right)^{1/2},$$

$$\sigma = \frac{p_c}{\rho_c c^2} = \frac{K\rho_c^{1/n}}{c^2}$$

(17)

$$\upsilon(\xi) = \frac{m}{M} = \frac{ZA^3 m}{8\pi\rho_c} = \frac{G\,m\,A}{\sigma(1+n)c^2}.$$

(18)

By rearranging terms, the FTOV equations will have the form

$$\xi^{2\alpha} D_\xi^\alpha \theta - Z\sigma(n+1)\xi^\alpha \upsilon D_\xi^\alpha \theta + \upsilon + \upsilon\sigma\theta + \sigma\xi^\alpha \theta D_\xi^\alpha \upsilon + \sigma^2 \xi^\alpha \theta^2 D_\xi^\alpha \upsilon = 0,$$

(19)

and

$$D_\xi^\alpha \upsilon = \xi^2 \theta^n,$$

(20)

with the initial conditions

$$\theta(0)=1 \quad ,\upsilon(0)=0 \quad ,D_\xi^{\alpha\alpha}\theta = D_\xi^\alpha\left(D_\xi^\alpha \theta\right).$$

To solve the FTOV equations could be written as

$$\xi^{2\alpha} D_\xi^\alpha \theta - Z\sigma(n+1)\xi^\alpha \upsilon D_\xi^\alpha \theta + \upsilon + \upsilon\,\sigma\,\theta + \sigma\xi^\alpha \theta D_\xi^\alpha \upsilon + \sigma^2 \xi^\alpha \theta^2 D_\xi^\alpha \upsilon = 0,$$

(21)

and

$$D_\xi^\alpha \upsilon = \xi^{2\alpha} \theta^n,$$

(22)

subject to the initial condition

$$\theta(0)=1 \quad ,\upsilon(0)=0 \quad ,D_\xi^{\alpha\alpha}\theta = D_\xi^\alpha\left(D_\xi^\alpha \theta\right).$$

(23)

Considering a series expansion in the form

$$\theta(\xi) = \sum_{m=0}^{\infty} A_m \xi^{2m},$$

(24)

we can rewrite Equation (24) in the form of fractional calculus as

$$\theta(\xi^\alpha) = A_0 + A_2 \xi^{2\alpha} + A_4 \xi^{4\alpha} + A_3 \xi^{6\alpha} + \ldots$$

(25)

where $\theta(0) = A_0 = 1$, then

$$\theta\left(\xi^{\alpha}\right)=1+\sum_{m=1}^{\infty}A_{m}\xi^{2\alpha m} \ . \tag{26}$$

By applying Jumarie's mRL derivative, Equation (A4) to Equation (26) yields, after some manipulations, the coefficients of the series expansion can be calculated using the two recurrence relations [31]

$$A_{m+1}=\frac{\sigma}{U_{m+1}}\left(Z(n+1)\gamma_{m-1}-\eta_{m}-\beta_{m}-\sigma^{2}\zeta_{m}\right)-\frac{N_{m}}{U_{m+1}}, \tag{27}$$

where

$$\zeta_{m}=\sum_{i=0}^{m}A_{i}\beta_{m-i},\ \beta_{m}=\sum_{i=0}^{m}A_{i}Q_{m-i},\ \eta_{m}=\sum_{i=0}^{m}A_{i}N_{m-i},\ \gamma_{m-1}=\sum_{i=0}^{m}f_{i}N_{m-i},\ f_{i}=A_{i+1}U_{i+1},$$

$$N_{m}=\frac{\Gamma(2\alpha m+2\alpha+1)}{\Gamma(2\alpha m+3\alpha+1)}Q_{2m},\ U_{m}=\frac{\Gamma(2\alpha m+1)}{\Gamma(2\alpha m-\alpha+1)},$$

and

$$Q_{m}=\frac{1}{\Gamma(m\alpha+1)A_{0}}\sum_{i=1}^{m}\frac{(m-1)!\Gamma(\alpha(m-i)+1)\Gamma(i\alpha+1)}{i!(m-i)!}(in-m+i)A_{i}Q_{m-i}\ \ \forall m\geq 1. \tag{28}$$

We applied the [33-34] approach, which combines the two accelerating techniques, the Euler-Abel transformation and the Pade approximation, to increase the series' convergence radius.

The radius $R$ and the mass $M(R)$ could be calculated from

$$R=A^{-1}\xi_{1}=\left(\frac{Zc^{2}}{8\pi G}(n+1)\sigma^{1-n}\left(\frac{K}{c^{2}}\right)^{n}\right)^{(1/2)}\xi_{1}, \tag{29}$$

and

$$M=\frac{4\pi\rho_{c}}{A^{3}}\upsilon(\xi_{1})=\left(\frac{1}{4\pi}\left(\frac{(n+1)c^{2}}{G}\right)^{3}\left(\frac{K}{c^{2}}\right)^{n}\right)^{(1/2)}\tilde{M}, \tag{30}$$

where

$$\tilde{M}=\left(\frac{Z}{2}\right)^{(1/2)}\sigma^{(3-n)/2}\upsilon(\xi_{1}), \tag{31}$$

$\xi_1$ is the first zero of the Lane-Emden function $\theta(\xi)$.

## 3. Numerical Results

The stability of fractional relativistic polytropes is analyzed using Equation (31), considering various values of the general relativity effect $\sigma$, mass $\tilde{M}(\sigma)$, and polytropic index $n$. The numerical results are meticulously tabulated for polytropic indices $n =1$, 1.5, 2, and 3 across a specified range of $\sigma$. Tables 1 to 4 focus on the critical values, denoted as $\sigma_{CR}$, corresponding to $\tilde{M}(\sigma)$ influenced by relativistic effects. These tables also highlight variations across different polytropic indices $n$ and several values of the fractional parameter $\alpha$, offering a comprehensive view of the relationships between these variables in the context of general relativity. Tables 1 to 4 show the critical values $\sigma_{CR}$ corresponding to $\tilde{M}(\sigma)$ due to relativistic effects for different polytropic indices and several values of the fractional parameter $\alpha$.

Table 1. The critical values of the relativistic parameters $\sigma_{CR}$ and the corresponding $\tilde{M}(\sigma)$ for $n = 1$.

| $\alpha$ | $\sigma_{CR}$ | $\xi_{1A}$ | $\upsilon(\xi)$ | $\tilde{M}(\sigma)$ |
|---|---|---|---|---|
| 1 | 0.42 | 1.888 | 0.595477 | 0.2501 |
| 0.99 | 0.45 | 1.867 | 0.55311 | 0.247141 |
| 0.98 | 0.45 | 1.882 | 0.550866 | 0.244411 |
| 0.97 | 0.47 | 1.878 | 0.525562 | 0.241849 |
| 0.96 | 0.47 | 1.897 | 0.523787 | 0.239364 |
| 0.95 | 0.47 | 1.917 | 0.523831 | 0.237738 |

Table 2. The critical values of the relativistic parameters $\sigma_{CR}$ and the corresponding $\tilde{M}(\sigma)$ for $n = 1.5$.

| $\alpha$ | $\sigma_{CR}$ | $\xi_{1A}$ | $\upsilon(\xi)$ | $\tilde{M}(\sigma)$ |
|---|---|---|---|---|
| 1 | 0.2 | 2.696 | 0.96423 | 0.288375 |
| 0.99 | 0.2 | 2.710 | 0.97041 | 0.288171 |
| 0.98 | 0.2 | 2.725 | 0.96268 | 0.283868 |
| 0.97 | 0.22 | 2.696 | 0.896795 | 0.282055 |
| 0.96 | 0.22 | 2.711 | 0.901219 | 0.281484 |
| 0.95 | 0.22 | 2.731 | 0.909952 | 0.282257 |

Table 3. The critical values of the relativistic parameters $\sigma_{CR}$ and the corresponding $\tilde{M}(\sigma)$ for $n = 2$.

| $\alpha$ | $\sigma_{CR}$ | $\xi_{1A}$ | $\upsilon(\xi)$ | $\tilde{M}(\sigma)$ |
|---|---|---|---|---|
| 1 | 0.1 | 3.698 | 1.2957 | 0.409757 |
| 0.99 | 0.1 | 3.691 | 1.3084 | 0.410843 |
| 0.98 | 0.1 | 3.713 | 1.3166 | 0.410522 |
| 0.97 | 0.1 | 3.751 | 1.3260 | 0.410579 |
| 0.96 | 0.1 | 3.763 | 1.33488 | 0.41044 |
| 0.95 | 0.1 | 3.788 | 1.3477 | 0.411546 |

Table 4. The critical values of the relativistic parameters $\sigma_{CR}$ and the corresponding $\tilde{M}(\sigma)$ for $n = 3$.

| $\alpha$ | $\sigma_{CR}$ | $\xi_{1A}$ | $\upsilon(\xi)$ | $\tilde{M}(\sigma)$ |
|---|---|---|---|---|
| 1 | 0 | 6.894 | 2.02059 | 2.02059 |
| 0.99 | 0 | 6.658 | 2.111623 | 2.11466 |
| 0.98 | 0.001 | 6.863 | 2.15300 | 2.12279 |
| 0.97 | 0.001 | 6.929 | 2.20077 | 2.15476 |
| 0.96 | 0.001 | 6.943 | 2.23836 | 2.17638 |
| 0.95 | 0.001 | 6.909 | 2.26399 | 2.18618 |

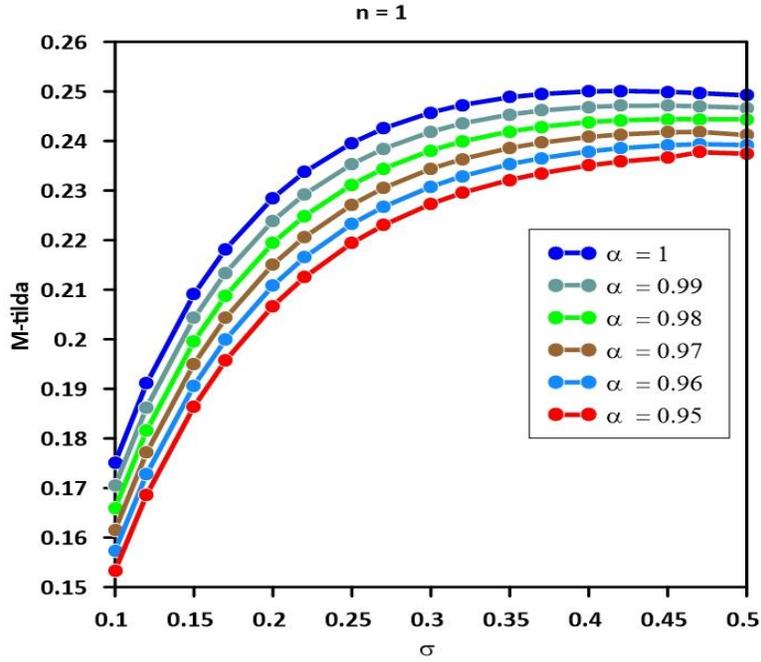

Figure 1. Stellar mass $\tilde{M}(\sigma)$ for polytropic index $n = 1$ across various values of the fractional parameter $\alpha$, with $\sigma_{CR}$ = 0.42 for $\alpha = 1$. Stability is observed for $\sigma < 0.42$; instability occurs when $\sigma > 0.42$.

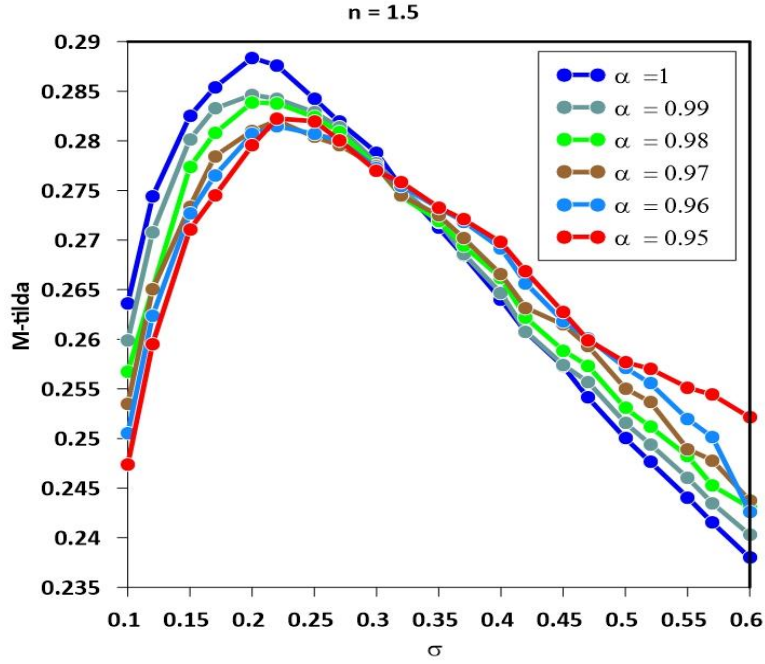

Figure 2. Stellar mass $\widetilde{M}(\sigma)$ for polytropic index $n=1.5$ across various values of the fractional parameter $\alpha$, with $\sigma_{CR}=0.2$ for $\alpha=1$. Stability is observed for $\sigma<0.2$; instability occurs when $\sigma>0.2$.

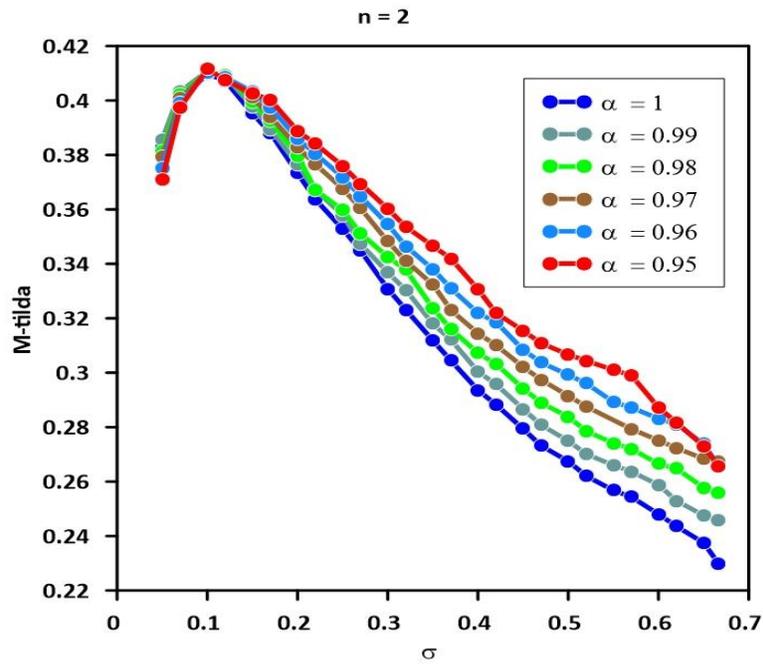

Figure 3. Stellar mass $\widetilde{M}(\sigma)$ for polytropic index $n=2$ across various values of the fractional parameter $\alpha$, with $\sigma_{CR}=0.1$ for $\alpha=1$. Stability is observed for $\sigma<0.1$; instability occurs when $\sigma>0.1$.

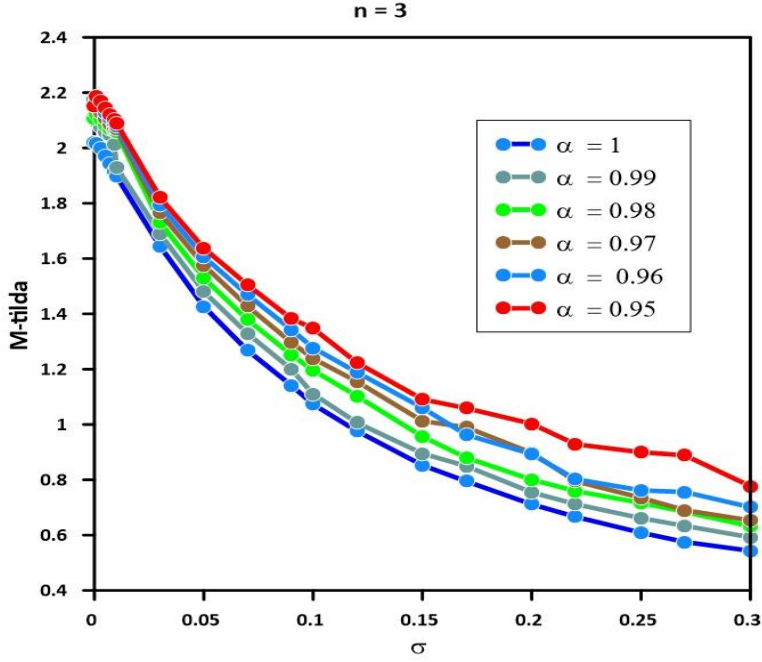

Figure 4. Stellar mass $\tilde{M}(\sigma)$ for polytropic index $n = 3$ across various values of the fractional parameter $\alpha$, with $\sigma_{CR} = 0$ for $\alpha = 1$. Stability is observed for $\sigma < 0$; instability occurs when $\sigma > 0$.

Figures 1 to 4 display $\tilde{M}$ from Equation (31) as a function of the polytropic index and the relativistic effect denoted as $\sigma$. These figures demonstrate a trend where $\tilde{M}$ (and consequently, the stellar mass) increases with $\sigma$ up to certain peak values, identified as $\sigma_{CR}$. It's important to note that the critical value $\sigma_{CR}$ indicates the commencement of the first mode of radial instability. For polytropic indices $n = (1, 1.5, 2, 3)$, the data shows that both the critical values and the stability of the relativistic polytrope change in response to variations in the fractional parameter $\alpha$. This relationship highlights the influence of $\alpha$ on the stability dynamics of the relativistic polytrope across different polytropic indices. To declare the effect of instability on the physical quantities of the star, we can take n=1.5 (Figure 2) as an example. For fixed $\alpha$ and changing $\sigma$, $\tilde{M}$ increases to a specific value of $\sigma$ represents the critical value $\sigma_{CR}$ and then decreases; this implies from Equation (31) that the mass function and, consequently, the star's mass will decrease. Another feature could be noticed from these tables (for example, Tabel 1 for n=1): as the fractional parameters decrease, the critical relativistic parameter increases, which indicates the influence of the fractional derivative and the internal structure of the star. From the definition of the relativistic parameter, $\sigma = k\rho_c^{1/n}/c^2$, the central density increases as the fractional parameter decreases.

Exploring the stability of polytropes is instrumental in identifying physical properties like the maximum mass limit, and it sheds light on variations in stellar mass resulting from general relativity's impact. The mass-radius relation [2] is expressed by the formula (see Appendix B for details)

$$\frac{GM}{c^2\bar{R}} = \frac{\sigma(n+1)\upsilon(\xi_1)}{\bar{\xi}_1}. \tag{32}$$

The mass-radius relationship is crucial for assessing surface redshift, as it provides the ratio of the gravitational radius $ZGM/c^2$ to the invariant radius $\bar{R}$, given known values of $n$ and $\sigma$. To apply this concept practically, Equation (32) is reformulated using the solar mass and radius as numerical values, followed by taking logarithms of both sides of this equation, as elaborated in [2].

Utilizing the solutions discussed in Section 2, Figures 5 to 10 offer insights into the mass-radius relation, which is integral to understanding the internal structure of a polytrope. This indicates that specific values of the relativistic parameter $\sigma$ correspond to distinct internal structures. To illustrate this relationship, we have graphed the logarithmic ratio of the gravitational radius to the geometrical radius as a function of $\sigma$, for various polytropic indices $n$ and fractional parameters $\alpha$, thereby mapping out the structural variations inherent to different values of $\sigma$.

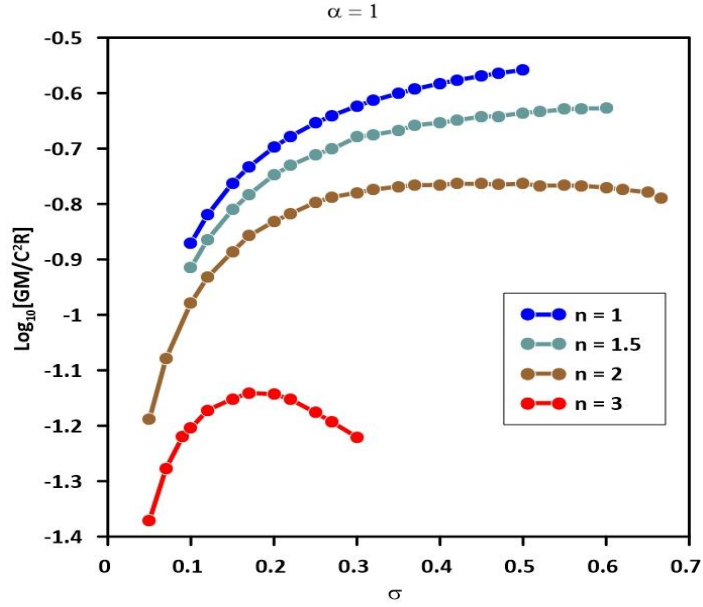

Figure 5. Illustration of the logarithmic ratio of the gravitational radius to the coordinate radius for a fractional parameter $\alpha = 1$, mapped as a function of the relativistic parameter $\sigma$, across various polytropic indices $n$.

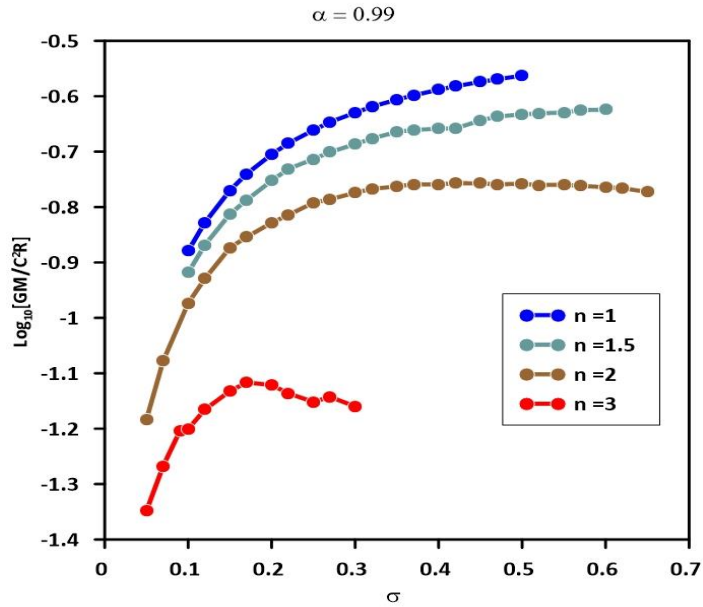

Figure 6. Illustration of the logarithmic ratio of the gravitational radius to the coordinate radius for a fractional parameter $\alpha = 0.99$, mapped as a function of the relativistic parameter $\sigma$, across various polytropic indices $n$.

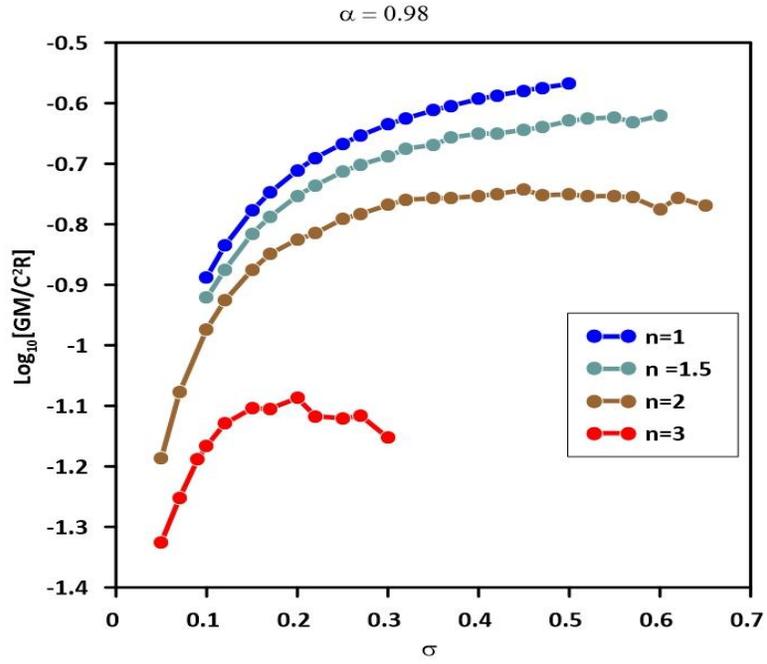

Figure 7. Illustration of the logarithmic ratio of the gravitational radius to the coordinate radius for a fractional parameter $\alpha = 0.98$, mapped as a function of the relativistic parameter $\sigma$, across various polytropic indices *n*.

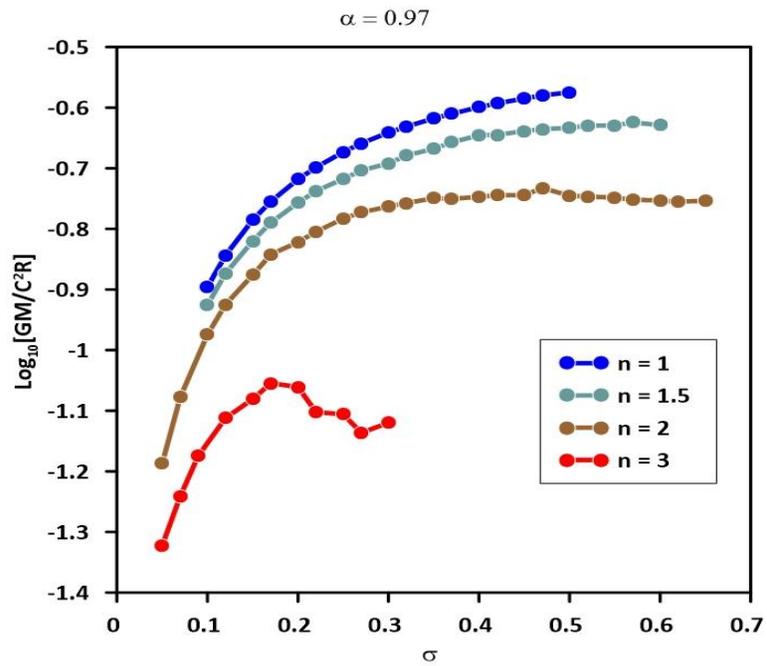

Figure 8. Illustration of the logarithmic ratio of the gravitational radius to the coordinate radius for a fractional parameter $\alpha = 0.97$, mapped as a function of the relativistic parameter $\sigma$, across various polytropic indices *n*.

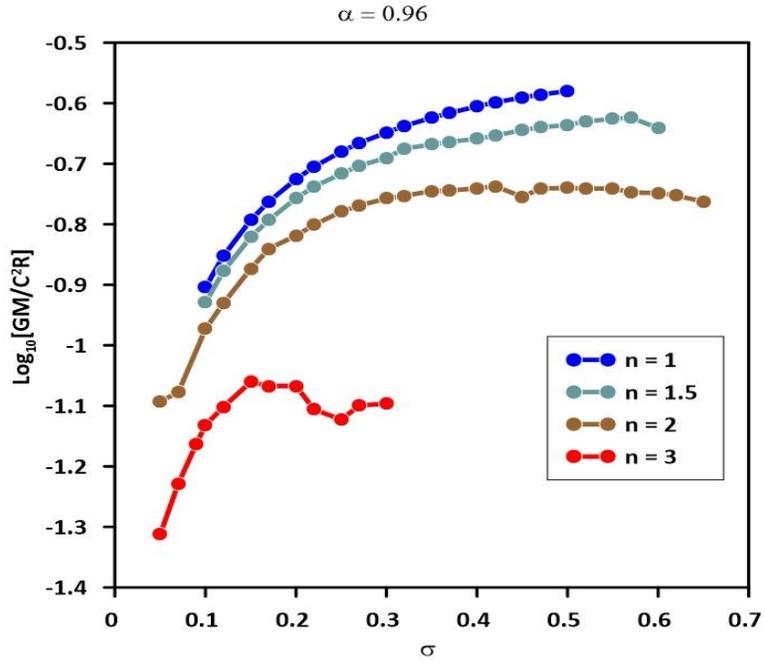

Figure 9. Illustration of the logarithmic ratio of the gravitational radius to the coordinate radius for a fractional parameter $\alpha = 0.96$, mapped as a function of the relativistic parameter $\sigma$, across various polytropic indices $n$.

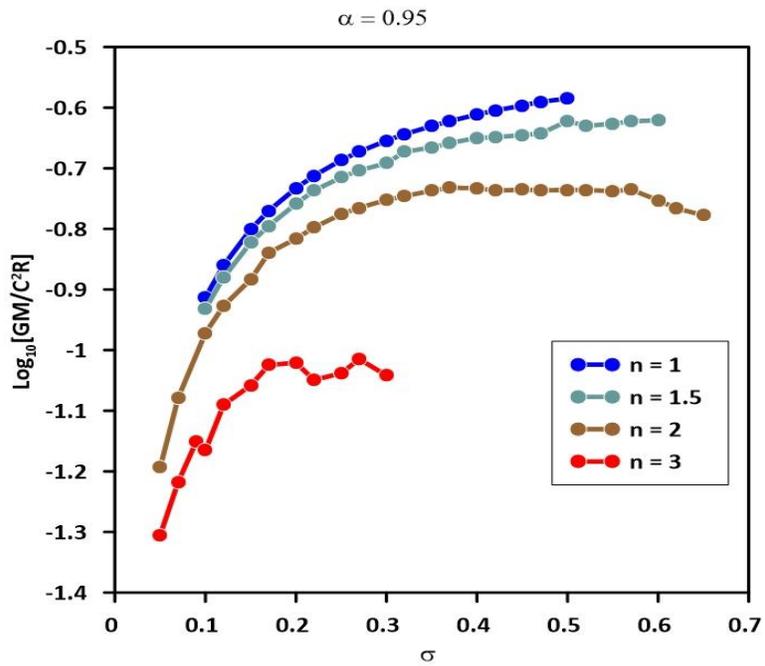

Figure 10. Illustration of the logarithmic ratio of the gravitational radius to the coordinate radius for a fractional parameter $\alpha = 0.95$, mapped as a function of the relativistic parameter $\sigma$, across various polytropic indices $n$.

The figures collectively show how the logarithmic ratio of the gravitational radius to the coordinate radius varies with changes in the relativistic parameter for different polytropic indices. These variations are crucial for understanding the internal structure of a polytrope and how relativistic effects influence it. Tables 5 to 10 give the limits of mass-radius relations for different polytropic indices $n$ with several values of fractional parameters.

The stability limit, observed at maximum mass within self-gravitating polytropic fluid spheres, arises from the delicate balance between gravitational attraction and internal pressure forces. This equilibrium is essential for maintaining hydrostatic equilibrium, which is a prerequisite for system stability [14, 35-37]. As the sphere's mass increases, so does the gravitational force pulling inward. Simultaneously, the pressure generated by the fluid within the sphere also rises because of the increased weight of the material above. However, there is a critical juncture where the gravitational force becomes sufficiently strong to overcome the pressure, prompting the collapse of the sphere. This collapse can induce further gravitational contraction, elevating the density and temperature within the sphere and potentially triggering processes like nuclear fusion in stars.

Table 5. Limits of mass-radius relations for $\alpha = 1$.

| $n$ | $\sigma$ | $\xi$ | $\bar{\xi}_1$ | $Log\ (\sigma(n+1)\upsilon(\xi_1)/\bar{\xi}_1)$ | Limit ratio of $GM/c^2\bar{R}$ | Limit ratio of $GM/c^2R$ |
|---|---|---|---|---|---|---|
| 1 | 0.5 | 1.801 | 2.31773 | -0.667485 | 0.215038 | 0.276735 |
| 1.5 | 0.6 | 2.219 | 3.03701 | -0.763529 | 0.172374 | 0.235917 |
| 2 | 0.45 | 3.261 | 4.30403 | -0.883938 | 0.130636 | 0.17242 |
| 3 | 0.17 | 7.477 | 8.63918 | -1.20371 | 0.0625594 | 0.0722833 |

Table 6. Limits of mass-radius relations for $\alpha = 0.99$.

| $n$ | $\sigma$ | $\xi$ | $\bar{\xi}_1$ | $Log\ (\sigma(n+1)\upsilon(\xi_1)/\bar{\xi}_1)$ | Limit ratio of $GM/c^2\bar{R}$ | Limit ratio of $GM/c^2R$ |
|---|---|---|---|---|---|---|
| 1 | 0.5 | 1.816 | 2.32136 | -0.669473 | 0.214056 | 0.273623 |
| 1.5 | 0.6 | 2.240 | 3.06528 | -0.760273 | 0.173671 | 0.237656 |
| 2 | 0.42 | 3.308 | 4.35031 | -0.875803 | 0.133106 | 0.175046 |
| 3 | 0.17 | 7.612 | 8.84601 | -1.18182 | 0.0657938 | 0.0764599 |

Table 7. Limits of mass-radius relations for $\alpha = 0.98$.

| $n$ | $\sigma$ | $\xi$ | $\bar{\xi}_1$ | $Log\ (\sigma(n+1)\upsilon(\xi_1)/\bar{\xi}_1)$ | Limit ratio of $GM/c^2\bar{R}$ | Limit ratio of $GM/c^2 R$ |
|---|---|---|---|---|---|---|
| 1 | 0.5 | 1.833 | 2.34465 | -0.674947 | 0.211375 | 0.270376 |
| 1.5 | 0.6 | 2.264 | 3.08377 | -0.754882 | 0.17584 | 0.23951 |
| 2 | 0.45 | 3.322 | 4.65552 | -0.889445 | 0.12899 | 0.180769 |
| 3 | 0.2 | 7.941 | 9.32122 | -1.1568 | 0.0696947 | 0.0818083 |

Table 8. Limits of mass-radius relations for $\alpha = 0.97$.

| $n$ | $\sigma$ | $\xi$ | $\bar{\xi}_1$ | $Log\ (\sigma(n+1)\upsilon(\xi_1)/\bar{\xi}_1)$ | Limit ratio of $GM/c^2\bar{R}$ | Limit ratio of $GM/c^2 R$ |
|---|---|---|---|---|---|---|
| 1 | 0.5 | 1.850 | 2.36174 | -0.680674 | 0.208606 | 0.26631 |
| 1.5 | 0.57 | 2.314 | 3.24426 | -0.770941 | 0.169457 | 0.237581 |
| 2 | 0.47 | 3.471 | 4.33571 | -0.829169 | 0.148194 | 0.185113 |
| 3 | 0.17 | 7.797 | 9.07025 | -1.12032 | 0.0758025 | 0.0881811 |

Table 9. Limits of mass-radius relations for $\alpha = 0.96$.

| $n$ | $\sigma$ | $\xi$ | $\bar{\xi}_1$ | $Log\ (\sigma(n+1)\upsilon(\xi_1)/\bar{\xi}_1)$ | Limit ratio of $GM/c^2\bar{R}$ | Limit ratio of $GM/c^2 R$ |
|---|---|---|---|---|---|---|
| 1 | 0.5 | 1.870 | 2.36711 | -0.682232 | 0.207859 | 0.263115 |
| 1.5 | 0.57 | 2.351 | 3.26239 | -0.766273 | 0.171288 | 0.23769 |
| 2 | 0.5 | 3.588 | 4.53849 | -0.842131 | 0.143836 | 0.18194 |
| 3 | 0.2 | 8.575 | 10.013 | -1.1344 | 0.0733841 | 0.0856899 |

Table 10. Limits of mass-radius relations for $\alpha = 0.95$.

| $n$ | $\sigma$ | $\xi$ | $\bar{\xi}_1$ | $Log\ (\sigma(n+1)\upsilon(\xi_1)/\bar{\xi}_1)$ | Limit ratio of $GM/c^2\bar{R}$ | Limit ratio of $GM/c^2 R$ |
|---|---|---|---|---|---|---|
| 1 | 0.5 | 1.891 | 2.45497 | -0.698346 | 0.200287 | 0.260021 |
| 1.5 | 0.6 | 2.399 | 3.33356 | -0.763731 | 0.172293 | 0.239412 |
| 2 | 0.37 | 3.487 | 4.5882 | -0.851331 | 0.140821 | 0.185293 |
| 3 | 0.27 | 10.284 | 12.0304 | -1.08293 | 0.0826169 | 0.0966468 |

## 4. Conclusions

Applying fractional derivatives in solving FTOV equations significantly advances our comprehension of compact astrophysical objects. This research underscores the importance of the critical relativistic parameter, $\sigma_{CR}$, at which the mass of a polytrope reaches its maximum value, indicating the beginning of radial instability. For a fractional parameter $\alpha = 1$, we identify critical values $\sigma_{CR}$ = 0.42, 0.20, 0.10, and 0.0 for polytropic indices $n = 1$, 1.5, 2, and 3, respectively, aligning with [16] and [21] findings. Decreasing values of $\alpha$, such as 0.99, 0.98, and 0.97, correspond to the increase in the critical relativistic parameter $\sigma_{CR}$. This reveals that both the critical values and stability of the relativistic polytrope are influenced by changes in the fractional parameter α, as depicted in Figures 1 to 4, demonstrating the impact of $\alpha$ on the stability dynamics of relativistic polytropes across different indices.

Another key discovery relates to the mass-radius relationship. For $\alpha = 1$ and polytropic index $n = 1.0, GM/c^2\bar{R} \leq 0.215$, meaning the gravitational radius $2GM/c^2$ is up to 43.0% of the physical (invariant) radius, while for $n = 3.0$, it is up to 14.4% of the radius, a notable decrease compared to the limit at $n = 1.0$. These findings, detailed in Tables 5 to 10 and illustrated in Figures 5 to 8, agree with [2] and [21] studies. The mass-radius relationship in relativistic polytropes is investigated, revealing a variety of correlations between the gravitational and physical radii of these polytropes for a range of relativistic parameters σ and polytropic indices n. All results are consistent with the findings of several other researchers.

The fractional derivatives approach offers profound insights into the behavior and characteristics of fractional polytropes within a relativistic framework. For instance with $\alpha = 0.99, 0.98, 0.97$, etc., diverse relationships emerge between the gravitational and physical radii of the relativistic polytrope across various relativistic parameters $\sigma$ and polytropic indices $n$. To illustrate the impact of instability on the physical properties of the star, we consider the case when n=1.5 (as shown in Figure 2). At fixed $\alpha$ and changing $\sigma$, $\tilde{M}$ increases to a specific value of $\sigma$ represents the critical value $\sigma_{cr}$ and then decreases; this implies from Equation (31) that the mass function and, consequently, the mass of the star will decrease.

Furthermore, we observed that when the fractional parameter decreases, the key relativistic parameter increases. This indicates that the effect of the fractional derivative on the internal structure of the star is a significant factor. Considering the definition of the relativistic parameter, it can be inferred that when the fractional parameter decreases, the central density increases.

**Acknowledgment:** The authors thank the editors and reviewers for their valuable comments.

**Data availability statement**

All data that support the findings of this study are included within the article (and any supplementary files).

**Statements and Declarations**

The authors declare that there is no conflict of interest.

**Competing interest**

The authors declare that they have no known competing financial interests or personal relationships that could have appeared to influence the work reported in this paper.

**Funding**

The authors did not receive support from any organization for the submitted work.

**Appendix A: mRL Derivatives**

Suppose that $f: R \to R$ where $x \to f(x)$ is a continuous function and $h$ denotes a constant discretization span. The mRL derivative of the function $f$ is given by [38-44]

$$f^{(\alpha)}(x) = \lim_{h \to \infty} \frac{\Delta^{\alpha}\left[f(x) - f(0)\right]}{h^{\alpha}}, \qquad 0 < \alpha < 1 \tag{A1}$$

where

$$\Delta^{\alpha} f(x) = \sum_{k=0}^{\infty} (-1)^k \frac{\Gamma(\alpha+1)}{\Gamma(k+1)\Gamma(\alpha-k+1)} f\left[x + (\alpha-k)h\right], \tag{A2}$$

where $\Gamma(x)$ represents the Gamma function.

This fractional derivative is comparable to the conventional derivatives rule, according to which a constant's order derivative, $\alpha$, is equal to zero.

$$D^\alpha f(x) = \begin{cases} \dfrac{1}{\Gamma(-\alpha)} \displaystyle\int_0^x (x-\xi)^{-\alpha-1}\left[f(\xi)-f(0)\right]d\xi & \alpha < 0 \\[1em] \dfrac{1}{\Gamma(1-\alpha)} \dfrac{d}{dx}\displaystyle\int_0^x (x-\xi)^{-\alpha}\left[f(\xi)-f(0)\right]d\xi & 0 < \alpha < 1 \\[1em] \dfrac{1}{\Gamma(n-\alpha)} \dfrac{d^n}{dx^n}\displaystyle\int_0^x (x-\xi)^{n-\alpha-1}\left[f(\xi)-f(0)\right]d\xi & n < \alpha < n+1,\ n \geq 1 \end{cases} \quad (A3)$$

A few helpful formulas that Jumarie changed may be summed up as

$$D_x^\alpha x^\gamma = \frac{\Gamma(\gamma+1)}{\Gamma(\gamma-\alpha+1)} x^{\gamma-\alpha}, \qquad \gamma > 0 \tag{A4}$$

$$D_x^\alpha (cf(x)) = c D_x^\alpha f(x), \tag{A5}$$

$$D_x^\alpha [f(x)g(x)] = g(x) D_x^\alpha f(x) + f(x) D_x^\alpha g(x), \tag{A6}$$

$$D_x^\alpha f[g(x)] = f_g'[g[x]] D_x^\alpha g(x), \tag{A7}$$

$$D_x^\alpha f[g(x)] = D_g^\alpha f[g[x]] (g_x')^\alpha, \tag{A8}$$

where $c$ is a constant.

Equations (A6) through (A8) follow directly from

$$D_x^\alpha f(x) \cong \Gamma(\alpha+1) D_x f(x), \tag{A9}$$

Following [43], equation (A7) could be modified to

$$D_x^\alpha f[g(x)] = \omega_x f_g'[g[x]] D_x^\alpha g(x). \tag{A10}$$

As a result, Equations (A6) and (A8) will be changed to the subsequent formats

$$D_x^\alpha [f(x)g(x)] = \omega_x \{g(x) D_x^\alpha f(x) + f(x) D_x^\alpha g(x)\}, \tag{A11}$$

$$D_x^\alpha f[g(x)] = \omega_x D_g^\alpha f[g[x]] (g_x')^\alpha, \tag{A12}$$

where $\omega_x$ is called the fractal index determined in terms of gamma functions as follows [42]

$$\omega_x = \frac{\Gamma(m\alpha+1)}{m\Gamma(m\alpha+1)\Gamma(m\alpha-\alpha+1)}. \tag{A13}$$

**Appendix B: Derivation of Equation (32)**

By using $\xi = Ar$ in Equation (18), we get

$$\upsilon(\xi) = \frac{Gm}{c^2\sigma(n+1)}\frac{\xi}{r}, \tag{B1}$$

where $r = R$, $\xi = \xi_1$ at $m = M$. The mass-radius relation is given by

$$\frac{GM}{c^2 R} = \frac{\sigma(n+1)\upsilon(\xi_1)}{\xi_1}, \tag{B2}$$

The distance from the center of a point with radial coordinates is given by

$$d\bar{r} = \int_0^r e^{2\lambda/2} dr. \tag{B3}$$

From Equation (7) and Equation (B2), we get

$$e^{-2\lambda} = 1 - \frac{Z\sigma(n+1)c^2\upsilon(\xi)}{\xi}, \tag{B4}$$

by writing $\bar{\xi} = A\bar{r}$ and using the last equation, the physical radius (invariant radius) of the sphere and can be obtained in mRL by integrating the Equation

$$\bar{\xi}_1 = \int_0^{\xi} \left(1 - Z\sigma(n+1)\upsilon(\xi)/\xi\right)^{-1/2} d\xi. \tag{B5}$$

Since the value $\bar{\xi}_1 = A\bar{R}$ corresponds to the physical radius (invariant radius) $\bar{R}$ of the sphere.

The mass and invariant radius of the sphere are related and can be written as

$$\frac{GM}{c^2\bar{R}} = \frac{\sigma(n+1)\upsilon(\xi_1)}{\bar{\xi}_1} \ . \tag{B6}$$